\documentclass[aps,twocolumn,10pt,amsmath,amssymb,nofootinbib,superscriptaddress]{revtex4-2}

\usepackage{epsfig} 
\usepackage{amsmath} 
\usepackage{graphicx}
\usepackage{color}
\usepackage{float}
\usepackage{soul,xcolor}
\usepackage[font=scriptsize]{caption}
\usepackage{braket}
\usepackage{bbold}
\usepackage{subfig}
\usepackage{braket}
\usepackage{titlesec}
\usepackage{placeins}
\usepackage{hyperref}
\usepackage{caption}
\usepackage{lipsum}
\begin{document}
\title{ Steering in Neutrino Oscillations with Non-Standard Interaction }

\author{Lekhashri Konwar}
\email{konwar.3@iitj.ac.in (Corresponding author)}

\author{Bhavna Yadav}
\email{yadav.18@iitj.ac.in}
\affiliation{Indian Institute of Technology Jodhpur, Jodhpur 342030, India}

\begin{abstract}
In this study, we analyze the influence of Non-Standard Interaction (NSI) on steering in three-flavor neutrino oscillations, with a focus on the NO$\nu$A and DUNE experimental setups. DUNE, having a longer baseline, exhibits a more pronounced deviation towards NSI in steering compared to NO$\nu$A. Within the energy range where DUNE's maximum flux appears, the steering value for DUNE shows a $21\%$ deviation from the Standard Model (SM) to NSI for normal ordering (NO), while for inverted ordering (IO), the steering value increases by approximately $15\%$ relative to the SM. We conduct a comparative analysis of nonlocality, steering, and entanglement. Additionally, we express steering in terms of three-flavor neutrino oscillation probabilities and explore the relationship between steering inequality and concurrence.

\end{abstract}

\maketitle
		
\section{Introduction}
 The nonlocal nature of quantum correlations is inherently a fundamental feature of quantum mechanics. The nonlocal and unique feature of quantum mechanics has undergone extensive investigation in the recent past and constitutes a significant component of quantum information theory. Through the violation of a suitable set of inequalities, as demonstrated by Bell \cite{Bell:1964}, displays the incompatibility of quantum correlations with local hidden variable (LHV) theory \cite{Einstein:1935}, a concept initially explored by Einstein, Podolsky, and Rosen (EPR) in 1935. And it became evident that the predictions of quantum mechanics stand in contradiction to classical relations.
 
In the same year, Schrödinger \cite{Schrödinger:1935} introduced the concept of quantum steering as a way to formalize what EPR famously called “spooky action at a distance”. Quantum steering refers to the scenario where one party can alter the state of a distant party through local measurements. In 2007, Wiseman, Jones, and Doherty provided a definition of steering as a quantum phenomenon that arises when the predictions of quantum mechanics cannot be explained by a classical-quantum framework, where one party is assumed to send pre-determined states to another. Furthermore, observing quantum steering can be interpreted as detecting entanglement when one party performs uncharacterized measurements \cite{Wiseman:2007}. In this sense, steering lies between nonlocality \cite{Brunner:2014} and entanglement \cite{Horodecki:2009,Gühne:2009}. As a result, the study of quantum steering has offered new insights into the understanding of quantum inseparability, making it an increasingly popular area of research.

The question of neutrino mass remains one of the significant mysteries in particle physics. Compelling evidence has emerged suggesting that neutrinos possess nonzero masses. Although these masses are extremely small, they could still have a considerable impact on the mass density of the universe. The evidence supporting nonvanishing neutrino masses stems from the observed phenomenon of neutrino oscillations \cite{Bahcall:2004,Eguchi:2003,Ashie:2004,Michael:2006,DUNE:2015lol,Adamson:2016,Abe:2014}, where a neutrino of one flavor changes into another. Since the SM cannot account for nonvanishing neutrino masses, this phenomenon points to physics beyond the SM.

NSI provides a promising extension to the SM in the context of neutrino. As neutrinos traverse the Earth's crust, they are influenced by the SM matter potential, commonly referred to as the matter effect. Beyond this, neutrinos may also experience NSI, which deviates from standard neutrino interactions with matter. These deviations are parameterized through effective couplings, denoted as $\epsilon_{\alpha\beta}$. These parameters can be real or complex, encapsulate potential new physics effects beyond the SM. Neutrino oscillation probabilities can be expressed in terms of these parameters in the presence of NSI, allowing their effects to be probed in neutrino experiments. Long-baseline neutrino experiments, with their sensitivity to matter effects, serve as an excellent platform for probing and constraining the influence of NSI.

Quantum correlations, initially studied in optical and electronic systems \cite{Aspect:1981, Tittel:1998}, have recently been extended to high-energy physics. Quantum correlations in the context of two-flavor and three-flavor neutrino oscillations have been explored in \cite{Blasone:2007wp,Blasone:2007vw,Alok:2014gya,Banerjee:2015mha,Formaggio_2016,Fu:2017hky,Naikoo:2017fos,Naikoo:2019,Dixit:2019swl,Shafaq:2020sqo,Sarkar:2020vob,Ming:2020nyc,Blasone:2021mbc,Yadav:2022grk,Chattopadhyay:2023xwr,konwar:12,Yadav:13,Rmohanta,Ming:2020,Siena:2021,Bittencourt:2022,Bittencourt:2024} where entanglement and nonlocality have been widely studied. However, steering \cite{paul:2020}, which lies between these two correlations, remains less explored. In this work, we aim to investigate steering in three-flavor neutrino oscillations by expressing it in terms of neutrino oscillation probabilities and analysing the effect of NSI.

The paper is organized as follows: Section \ref{sec2} outlines the general framework for including NSI effects in neutrino oscillation dynamics. Section \ref{sec3} provides an overview of the quantum steering measure. Next, in Section \ref{sec4} we explore steering within both two and three-flavor neutrino oscillation frameworks and express the steering inequality in terms of neutrino oscillation probabilities. Section \ref{sec5} presents our analysis of NSI impacts on steering in the context of the NO$\nu$A and DUNE experimental set-ups and discusses our results. Finally, Section \ref{sec6} summarizes the conclusions of the paper.

\section{Formalism}
\label{sec2}
\emph {Three flavor neutrino dynamics }: Consider a neutrino state $ \ket{\Psi\left ( t \right )}$ at time t, which can be written either in mass states $\left\{\nu _{1 }, \nu _{2}, \nu _{3}\right\}$ or in flavor states $\left\{\nu _{e }, \nu _{\mu  }, \nu _{\tau }\right\}$ as:
\begin{eqnarray}
    \ket{\Psi\left ( t \right )}= \sum_{\alpha = e,\mu ,\tau }^{}\nu _{\alpha }\left ( t \right ) \ket{\nu_{\alpha}}, \nonumber \\ or \nonumber \\
    = \sum_{i = 1, 2, 3 }^{}\nu _{i }\left ( t \right ) \ket{\nu_{i}}.
\end{eqnarray}
The flavor states $\nu_{e}$, $\nu_{\mu}$, and $\nu_{\tau}$ are expressed as a
combination of the mass eigenstates $\nu_{1}$, $\nu_{2}$, and $\nu_{3}$ and 
represented by the equation
\begin{equation}\label{2}
    \ket{\nu_{\alpha}}= \sum_{i}U_{\alpha i}\ket{\nu _{i}}, 
\end{equation}
and, vice-versa, we can also write
\begin{equation}
    \ket{\nu_{i}}= \sum_{\alpha}U^{*}_{\alpha i}\ket{\nu _{\alpha}}.
\end{equation}
The elements of the PMNS matrix are given by
\begin{equation}
\begin{pmatrix}
c_{12}c_{13} & s_{12}c_{13}&s_{13}e^{-\iota\delta}\\ 
-s_{12}c_{23}-c_{12}s_{13}s_{23}e^{\iota\delta}& c_{12}c_{23}-s_{12}s_{13}s_{23}e^{\iota\delta }&c_{13}s_{23}\\ 
s_{12}s_{23}-c_{12}s_{13}c_{23}e^{\iota\delta }& -c_{12}s_{23}-s_{12}s_{13}c_{23}e^{\iota\delta}&c_{13}c_{23}
\end{pmatrix}.
 \end{equation}
Here $s_{ij} = \sin \theta _{ij}$ and $c_{ij} = \cos \theta _{ij}$. These matrix elements are parameterized by mixing angles, such as $\theta_{12}$, $\theta_{13}$, and $\theta_{23}$, there exists a CP-violating phase, denoted as $\delta_{CP}$.\\
The evolution of the mass eigenstates $\ket{\nu_{i}(t)}$ is expressed as follows
\begin{equation}\label{4}
     \ket{\nu_{i}(t)}=e^{-\iota  E_{i}t}\ket{\nu_{i }}.
\end{equation}
Substituting the Eq. \eqref{2} into Eq. \eqref{4}, we can get the time evolution of flavor neutrino states $\ket{\nu_{\alpha}(t)}$ as
\begin{eqnarray}\label{5}
    \ket{\nu _{\alpha}(t)}=\sum_{i}  U_{\alpha i} e^{-\iota  E_{i}t} \ket{\nu_{i}} = U_{f}(t)\ket{\nu _{\alpha}},
\end{eqnarray}
considering, the state $\ket{\nu_{i}(t)}$ is the mass eigenstates of the free Dirac Hamiltonian $\mathcal{H}_{m}$ with energy $E_{i}$ $(i=1,2,3)$, then we can get the Eq. \eqref{5} as
\begin{eqnarray}\label{}
    \ket{\nu _{\alpha }(t)}=\Bar{U}_{\alpha e}(t)\ket{\nu}_{e}+\Bar{U}_{\alpha \mu}(t) \ket{\nu}_{\mu}\\ \nonumber
    +\Bar{U}_{\alpha \tau}(t) \ket{\nu}_{\tau}.
\end{eqnarray}
Defining $\Bar{U}=U e^{-i \mathcal{H}_m t} U^{\dagger}$ with $\mathcal{H}_{m} = \text{diag}(E_{1}, E_{2}, E_{3})$, where $E_{i}=\sqrt{m_{i}^{2}+p^{2}},$ $a=1,2,3$, assume the momentum p to be consistent across all mass eigenstates.

In the relativistic limit, neutrino flavor states are treated as distinct and individual modes within the framework of quantum field theory. The neutrino mode entanglement is that the time-evolved flavor state of neutrinos can be conceptualized as an entangled superposition of flavor modes. In essence, the distinct flavor states of neutrinos become intricately interconnected, such that the evolution of one mode influences the behaviour of others, and the time-evolved flavor state can be understood as an entangled superposition of the individual flavor modes, reflecting the inherent entanglement among them. As a consequence, the three-flavor neutrino system can be expressed  in flavor mode  as follows \cite{Blasone:2007vw}
\begin{eqnarray}\label{m4}
    \ket{\nu _{e}}\equiv \ket{1}_{e} \ket{0}_{\mu }\ket{0}_{\tau }\notag \\
    \ket{\nu_{\mu }}\equiv\ket{0}_{e } \ket{1}_{\mu } \ket{0}_{\tau }\\
   \ket{\nu_{\tau }}\equiv\ket{0}_{e } \ket{0}_{\mu } \ket{1}_{\tau }.\notag
\end{eqnarray}
\emph {Neutrino oscillations in matter}: As neutrinos travel through matter, their interaction with electrons introduces an additional term that significantly alters their behavior. This interaction occurs due to the weak force, which allows neutrinos to interact with electrons by exchanging W and Z bosons, and neutrinos can undergo charged current and neutral current. This consideration plays a crucial role in phenomena such as neutrino oscillations and is essential for understanding neutrino behavior in various experimental settings. 
\\
To incorporate the matter effect, the Hamiltonian of the SM can be expressed as follows \cite{Giunti:2007ry}
\begin{eqnarray}\label{osc3}
\mathcal{H}_m=\mathcal{H}_{vac}+\mathcal{H}_{mat}~~~~~~~~~~~~~~~~~~~~~~~ ~~~~~\nonumber\\
=\begin{pmatrix}
E_{1} & 0 & 0 \\ 
0 & E_{2} & 0 \\
0 & 0 & E_{3}
\end{pmatrix}+
U^\dagger\begin{pmatrix}
A & 0 & 0 \\ 
0 & 0 & 0 \\
0 & 0 & 0 
\end{pmatrix}U.
\end{eqnarray}
With $ A=\pm 2\sqrt{2}G_{F}N_{e},$ is the standard matter potential, due to the Coherent forward scattering of an electron neutrino from electrons in matter. $G_{F}$ is the Fermi constant and $N_{e}$ is the electron number density. For neutrinos, the sign of A is positive, while for antineutrinos, it is negative. The diagonal nature of this term implies that it affects different neutrino flavors independently, leading to distinct effects on their propagation. In the context of our analysis, the matter potential, denoted as 
$A$= $1.01  \times  10^{-13}$ eV, which is associated with an Earth’s matter density of 
$\rho= 2.8$ gm/cc.\\
\emph {Neutrino oscillations in Non-Standard Interaction (NSI)}:
The concept of NSIs has been introduced to account for sub-leading effects in neutrino flavor transitions, with future experiments potentially uncovering 'new physics' beyond the Standard Model in the form of unknown couplings involving neutrinos. \\
The widely studied dimension six four-fermion operators responsible for NSIs can be written as \cite{Wolfenstein:1978,Grossman:1995}
\begin{eqnarray}\label{nsi1}
\mathcal{L}_{NSI}^{NC}=2\sqrt{2}G_F\sum\limits_{\alpha, \beta, C} \epsilon_{\alpha\beta}^{f,C}(\bar{\nu}_\alpha \gamma^{\mu} P_{L} \nu_\beta)(\bar{f} \gamma_\mu P_{C} f),\nonumber \\
\end{eqnarray}
the parameters $\epsilon_{\alpha\beta}^{f,C}$
represent NSI effects in neutrino oscillations, where $\alpha$ and $\beta$ take on values e, $\mu$ and $\tau$, which correspond to the three neutrino flavors. f can be e (electron), u (up quark), or d (down quark), representing the types of particles with which neutrinos interact and C= L, R. \\
The total Hamiltonian in the presence of matter with NSI can be written as
\begin{eqnarray}\label{nsi2}
\mathcal{H}_{tot}&=&
\begin{pmatrix}
E_{1} & 0 & 0 \\ 
0 & E_{2}& 0 \\
0 & 0 & E_{3}

\end{pmatrix}\nonumber\\&&+U^{\dagger } A\begin{pmatrix}
1+\epsilon_{ee}(x) &\epsilon_{e\mu}(x)  &\epsilon_{e\tau}(x) \\ 
\epsilon_{\mu e}(x) & \epsilon_{\mu \mu}(x) & \epsilon_{\mu \tau}(x)\\
 \epsilon_{\tau e}(x) & \epsilon_{\tau \mu}(x) & \epsilon_{\tau \tau}(x)
\end{pmatrix} U.
\end{eqnarray}
Where \begin{equation}\label{nsi3}
    \epsilon _{\alpha \beta }(x)=\sum_{C,f=e,u,d}\frac{N_{f} (x) }{N_{e}(x)} \epsilon _{\alpha \beta }^{f},
\end{equation}
$\epsilon _{\alpha \beta }^{f}$ are the NSI parameters, which describe the interaction strengths between neutrinos of flavors $\alpha$ and $\beta$ with fermions of type f and $N_{f}$ is the number density of f type fermions.\\
In the context of three-flavor neutrino oscillations, the evolution operator $U_{f}(L)$ can be expressed as follows \cite{Ohlsson:1999xb}
\begin{eqnarray}\label{}
U_{f}(L)&=&\phi \sum_{a=1}^{3}e^{-i L\lambda _{a}}\frac{1}{3\lambda _{a}^2+c_{1}}\left [( \lambda _{a}^{2}+c_{1}) I+\lambda _{a}\tilde{T}+\tilde{T}^{2}\right ].\nonumber \\
\end{eqnarray}
Here $\phi = e^{iLTr\mathbf{H_{m}}/3}$,  $\tilde{T}=UTU^{-1}$ and $c_{1}=detT \times Tr T^{-1}$ and  $\lambda _{1}$, $\lambda _{2}$ and $\lambda _{3}$ are the eigenvalues of the matrix $T$, which is given as
\begin{equation}\label{osc5}
T \equiv \mathcal{H}_{tot}- (Tr H_{tot})I/3 =
\begin{pmatrix}
T_{11} & T_{12}& T_{13} \\
T_{21} & T_{22}& T_{23} \\
T_{31} & T_{32}& T_{33} 
\end{pmatrix}.
\end{equation}\\
The evolution operator $U_{f}(L)$ governs the change in the neutrino's flavor state over a distance L, incorporating the influences of NSI or matter effects. The oscillation probabilities, $\Tilde{ P}_{\alpha \beta }$, which give the probability that a neutrino of flavor $\alpha$ will be detected as flavor $\beta$ after traveling a distance L is expressed as:
\begin{equation}
   \Tilde{ P}_{\alpha \beta }\equiv \left|A_{\alpha \beta } \right|^{2}= \left| \bra\beta U_{f}(L) \ket\alpha\right|^{2}.
\end{equation}

\section{Steering }
\label{sec3}

Quantum steering \cite{uola:2020,canti:2016,Costa:2016,yang:2020} is a type of quantum correlation that is often described in the context of a quantum information task. 
Consider a scenario where two observers, Alice and Bob, are spatially separated and share a bipartite quantum state $\rho_{AB}$. They each have the ability to perform measurements from their respective sets, for Alice $A_{x}$  and for Bob $B_{y}$.  During a steering test, Bob seeks to determine whether the shared state is indeed entangled. To be convinced of the entanglement, Bob needs to confirm that the outcomes on his side are genuinely influenced by Alice's measurements, as opposed to being explainable by preexisting local hidden states (LHSs) that Alice might control. Thus, for Bob to validate the entanglement, he must effectively rule out the LHS model \cite{Wiseman:2007}
\begin{eqnarray}
     P\left ( a,b| A,B,\rho _{AB}\right )=\sum_{\lambda }^{}p_{\lambda }P\left ( a|A,\lambda  \right )P_{Q}\left ( b|B,\rho _{\lambda } \right ),\nonumber\\
\end{eqnarray}
here $P\left ( a,b| A,B,\rho _{AB}\right )=Tr\left ( A_{a}\otimes B_{b} \rho _{AB} \right )$
represents the probability of obtaining outcomes a and b when measurements A and B are performed on the shared state $\rho _{AB}$ by Alice and Bob, respectively. Here, $A_{a}$ and $B_{b}$ are their respective measurement operators. The hidden variable $\lambda$ corresponds to a state that Alice sends with probability $p_{\lambda }$ (where $\sum_{\lambda }^{} p_{\lambda }=1 )$. $P\left ( a|A,\lambda  \right )$ denotes the conditional probability of Alice obtaining outcome a under the hidden variable $\lambda$, while $P_{Q}\left ( b|B,\rho _{\lambda } \right )$ is the quantum probability of Bob obtaining outcome b when measuring B on the local hidden state $\rho_{\lambda}$. If Bob can show that the joint probability distribution $P\left ( a,b| A, B,\rho _{AB}\right )$ cannot be reproduced by any LHS theory, he will be convinced that Alice can steer his state, confirming that the shared bipartite state is entangled. Thus, a bipartite state $\rho_{AB}$ is unsteerable from Alice to Bob if and only if the joint probability distributions meet the criteria set by an LHS theory for all possible measurements A and B. The LHS theory assumption leads to specific steering inequalities, whose violation indicates the presence of steering.

A set of linear steering inequalities was formulated by Cavalcanti et al. \cite{caval:2009} to determine whether a bipartite state is steerable from Alice to Bob. These inequalities are relevant when both Alice and Bob can perform n dichotomic measurements on their respective subsystems
\begin{equation}\label{st1}
     F_{n}\left ( \rho _{AB} ,\mu \right )=\frac{1}{\sqrt{n}}\left|\sum_{k=1}^{n} \left< A_{k}\otimes B_{k}\right>\right|\leq 1,
\end{equation}
here $A_{k}=\hat{a}_{k} . \vec{\sigma }$ and $B_{k}=\hat{b}_{k} . \vec{\sigma }$, where $ \vec{\sigma }=( {\sigma_{1}}, {\sigma_{2}}, {\sigma_{3}})$ represents the Pauli matrices, and
$\hat{a}_{k}, \hat{b}_{k} \in  \mathbb{R}^{3}$ are unit vectors that are orthonormal. The term $\left< A_{k}\otimes B_{k}\right>$  is defined as $Tr \left [ \rho _{AB} (A_{k}\otimes B_{k}) \right ]$ , representing the expectation value when the bipartite state $\rho _{AB}$ is measured. The set $\mu =  \left\{ \hat{a}_{1}, \hat{a}_{2},...,\hat{a}_{n},\hat{b}_{1},\hat{b}_{2},....,\hat{b}_{n}\right\}$ includes the measurement directions chosen by Alice and Bob.\\
These inequalities serve as a criterion for identifying steerability, as their violation indicates that the correlations observed between Alice and Bob cannot be explained by any LHS theory.\\
In the Hilbert-Schmidt representation, any two-qubit state can be expressed as \cite{Luo:2008}
\begin{eqnarray}
    \rho_{AB} =\frac{1}{4} \left [I \otimes I  + \vec{a} .\vec{\sigma } \otimes I+ I \otimes \vec{b} .\vec{\sigma }+ \sum_{i,j}^{}t_{ij} \sigma _{i} \otimes \sigma _{j} \right ], \nonumber \\ 
\end{eqnarray}
where $ \vec{a}$ and $\vec{b}$ are the local Bloch vectors, $t_{ij}= Tr \left [ \rho _{AB} (\sigma_{i}\otimes \sigma_{j}) \right ]$, and $T_{AB}=\left [t_{ij}\right]$ is the correlation matrix. For the two measurement settings, state $\rho_{AB}$ is steerable if and only if \cite{paul:2020}
\begin{equation}
   S_{AB} = Tr\left [ T_{AB}^{T} T_{AB} \right ]> 1,
   \end{equation}
the notation T indicates the transpose of the correlation matrix $T_{AB}$. Among the three bipartite reduced states derived from a three-qubit state $\rho_{ABC}$, $S^{max}(\rho_{ABC} )$ is characterized by the maximum violation of the steering inequality \cite{paul:2020,Qui:2015,Dai:2022,Dong:2022,Dong:2023,qiu:2024}

\begin{equation}\label{sabc}
    S^{max}(\rho_{ABC})=max \left\{ S_{AB},S_{AC},S_{BC} \right\}.
\end{equation}

\section{Steering in terms of Neutrino Oscillations Probabilities}
\label{sec4}
\subsection{\emph{Two flavor:}}
In two-flavor, the neutrino state is expressed as:
\begin{eqnarray}
    \ket{\nu _{\alpha }\left ( t \right )}=\Bar{U}_{\alpha e}\left ( t \right )\ket{10}_{e}+\Bar{U}_{\alpha \mu}\left ( t \right )\ket{01}_{\mu}.
\end{eqnarray}

The density matrix for a two-flavor neutrino state denoted as $\rho _{AB}^{\alpha}(t) = \ket{\nu _{\alpha }(t)}\bra{\nu _{\alpha }(t)}$ encapsulates the evolution of a neutrino in a specific flavor state at time t. The matrix takes the form:

\begin{eqnarray}{\label{rho2}}
    \rho _{AB}^{\alpha}=\begin{pmatrix}
0 & 0 & 0 & 0 \\
0 & \rho _{11}^{\alpha} & \rho _{12}^{\alpha} & 0 \\
0 & \rho _{21}^{\alpha} & \rho _{22}^{\alpha} & 0 \\
0 & 0 & 0 & 0 \\
\end{pmatrix},
\end{eqnarray}
where the elements of this matrix can be written as
\begin{eqnarray}{\label{rhop}}
    \rho _{11}^{\alpha}= \left |\Bar{U}_{\alpha \mu }(t) \right |^{2};~~~~ \\ \rho _{12}^{\alpha}= \Bar{U}_{\alpha \mu }(t)\Bar{U}_{\alpha e }^{\ast }(t); \notag\\
      \rho _{21}^{\alpha}= \Bar{U}_{\alpha e }(t)\Bar{U}_{\alpha \mu }^{\ast }(t);\notag \\ \rho _{22}^{\alpha}= \left |\Bar{U}_{\alpha e }(t) \right |^{2}.~~~~~\notag
\end{eqnarray}
The associated probabilities for initial flavor, $\alpha =$ e are as follows: $P_{\alpha e}(t)= \left |\Bar{U}_{\alpha e }(t) \right |^{2}$ and $P_{\alpha \mu}(t)= \left |\Bar{U}_{\alpha \mu }(t) \right |^{2}$. The correlation matrix
\begin{eqnarray}
   T_{AB}=\left [t_{ij}\right],  t_{ij}= Tr \left [ \rho _{AB}^{\alpha} (\sigma_{i}\otimes \sigma_{j}) \right ]. 
\end{eqnarray}
The two-flavor steering measure $S_{AB}$, which is derived from the correlation matrix $T_{AB}$, is defined as:
\begin{eqnarray} \label{sab}
S_{AB} = Tr\left [ T_{AB}^{T} T_{AB} \right ]= \left\{ \rho _{11}^{\alpha} + \rho _{22}^{\alpha}\right\}^{2} + 8 \rho _{11}^{\alpha} \rho _{22}^{\alpha}.
\end{eqnarray}
In terms of neutrino oscillation probabilities can be written as \cite{Ming:2020}
\begin{eqnarray}\label{2f}
S_{AB}= \left\{P_{\alpha e}(t)+P_{\alpha \mu}(t)\right\}^{2} +8 P_{\alpha e}(t) P_{\alpha \mu}(t).
\end{eqnarray}

\subsection{\emph{Three flavor:}}
The time evolution for the neutrino quantum state into the flavor state can be written as
\begin{eqnarray}\label{23}
    \ket{\nu _{\alpha }\left ( t \right )}=\Bar{U}_{\alpha e}\left ( t \right )\ket{100}_{e}+\Bar{U}_{\alpha \mu}\left ( t \right )\ket{010}_{\mu} +\Bar{U}_{\alpha \tau}\left ( t \right ) \ket{001}_{\tau}. \nonumber \\
\end{eqnarray}
The density matrix $\rho _{ABC}^{\alpha}(t) = \ket{\nu _{\alpha }(t)}\bra{\nu _{\alpha }(t)}$ for neutrino is expressed as
\begin{equation}{\label{rho1}}
    \rho _{ABC}^{\alpha}(t)=\begin{pmatrix}
0 & 0 & 0 & 0 & 0 & 0 & 0 & 0\\ 
0 & \rho _{22}^{\alpha } & \rho _{23}^{\alpha } & 0 & \rho _{25}^{\alpha } & 0 & 0 & 0\\ 
0 & \rho _{32}^{\alpha } & \rho _{33}^{\alpha } & 0 & \rho _{35}^{\alpha } & 0 & 0 & 0\\ 
0 & 0 & 0 & 0 & 0 & 0 & 0 & 0\\ 
0 & \rho _{52}^{\alpha } & \rho _{53}^{\alpha } & 0 & \rho _{55}^{\alpha } & 0 & 0 & 0\\ 
0 & 0 & 0 & 0 & 0 &0  & 0 & 0\\ 
0 & 0 & 0 & 0 & 0 &0  & 0 & 0\\
0 & 0 & 0 & 0 & 0 &0  & 0 & 0 
\end{pmatrix},
\end{equation}
where the elements of this matrix can be represented as

\begin{eqnarray}{\label{rhop}}
    \rho _{22}^{\alpha}= \left |\Bar{U}_{\alpha \tau }(t) \right |^{2};~~~~~\notag \\ \rho _{23}^{\alpha}= \Bar{U}_{\alpha \tau }(t)\Bar{U}_{\alpha \mu }^{\ast }(t);\notag \\\rho _{25}^{\alpha}= \Bar{U}_{\alpha \tau }(t)\Bar{U}_{\alpha e }^{\ast }(t);\notag \\
    \rho _{32}^{\alpha}= \Bar{U}_{\alpha \mu }(t)\Bar{U}_{\alpha \tau }^{\ast }(t);\notag \\ \rho _{33}^{\alpha}= \left |\Bar{U}_{\alpha \mu }(t) \right |^{2};~~~~ \\ \rho _{35}^{\alpha}= \Bar{U}_{\alpha \mu }(t)\Bar{U}_{\alpha e }^{\ast }(t); \notag\\
     \rho _{52}^{\alpha}= \Bar{U}_{\alpha e }(t)\Bar{U}_{\alpha \tau }^{\ast }(t);\notag \\ \rho _{53}^{\alpha}= \Bar{U}_{\alpha e }(t)\Bar{U}_{\alpha \mu }^{\ast }(t);\notag \\ \rho _{55}^{\alpha}= \left |\Bar{U}_{\alpha e }(t) \right |^{2}.~~~~~\notag
\end{eqnarray}\\
The probabilities are as follows: $P_{\alpha e}(t)= \left |\Bar{U}_{\alpha e }(t) \right |^{2}$, $P_{\alpha \mu}(t)= \left |\Bar{U}_{\alpha \mu }(t) \right |^{2}$,  and $P_{\alpha \tau}(t)= \left |\Bar{U}_{\alpha \tau }(t) \right |^{2}$. As the neutrino survival and transition probabilities are dependent on the energy difference $\Delta E_{ij} = E_{i}-E_{j} \left ( i,j=1,2,3 \right )$. The energy difference $E_i -E _j$ can be approximated as $E_i -E j \approx \Delta m_{ij}^{2}/2 E$, where $E\equiv \left| \vec{P} \right|$, especially when this difference is much smaller than the neutrino energy E and length L is approximately equal to time t (i.e., $L\equiv t$) in the ultra-relativistic limit.\\
The reduced density matrices for neutrino $ \rho _{ABC}^{\alpha}(t)$ are
 \begin{eqnarray}
     \rho _{AB}^{\alpha}=Tr _{C}\left [\rho _{ABC}^{\alpha}  \right ] =\begin{pmatrix}
\rho _{22}^{\alpha} & 0 & 0 & 0 \\
0 & \rho _{33}^{\alpha} & \rho _{35}^{\alpha} & 0 \\
0 & \rho _{53}^{\alpha} & \rho _{55}^{\alpha} & 0 \\
0 & 0 & 0 & 0 \\
\end{pmatrix},
 \end{eqnarray}
 \begin{eqnarray}
     \rho _{AC}^{\alpha}=Tr _{B}\left [\rho _{ABC}^{\alpha}  \right ] =\begin{pmatrix}
\rho _{33}^{\alpha} & 0 & 0 & 0 \\
0 & \rho _{22}^{\alpha} & \rho _{25}^{\alpha} & 0 \\
0 & \rho _{52}^{\alpha} & \rho _{55}^{\alpha} & 0 \\
0 & 0 & 0 & 0 \\
\end{pmatrix},
 \end{eqnarray}
 \begin{eqnarray}
     \rho _{BC}^{\alpha}=Tr _{A}\left [\rho _{ABC}^{\alpha}  \right ] =\begin{pmatrix}
\rho _{55}^{\alpha} & 0 & 0 & 0 \\
0 & \rho _{22}^{\alpha} & \rho _{23}^{\alpha} & 0 \\
0 & \rho _{32}^{\alpha} & \rho _{33}^{\alpha} & 0 \\
0 & 0 & 0 & 0 \\
\end{pmatrix}.
 \end{eqnarray}
The $S_{AB}$, $S_{AC}$, and $S_{BC}$, as mentioned in Eq. \ref{sabc} are defined as follows:
\begin{equation} \label{Sab}
  S_{AB} =  Tr\left [ T_{AB}^{T} T_{AB} \right ],  
\end{equation}
\begin{equation} \label{Sac}
    S_{AC} =  Tr\left [ T_{AC}^{T} T_{AC} \right ],
\end{equation}
\begin{equation} \label{Sbc}
   S_{BC} =  Tr\left [ T_{BC}^{T} T_{BC} \right ] .
\end{equation}
Where $T_{AB}$, $T_{AC}$ and $T_{BC}$ are the correlation matrices of the density matrices $ \rho _{ABC}^{\alpha}(t)$ and the elements of these correlation matrices are defined by:
\begin{eqnarray}
   T_{AB}=\left [t_{ij}\right],  t_{ij}= Tr \left [ \rho _{AB}^{\alpha} (\sigma_{i}\otimes \sigma_{j}) \right ] , 
\end{eqnarray}

\begin{eqnarray}
   T_{AC}=\left [t_{ij}\right],  t_{ij}= Tr \left [ \rho _{AC}^{\alpha} (\sigma_{i}\otimes \sigma_{j}) \right ] , 
\end{eqnarray}

\begin{eqnarray}
   T_{BC}=\left [t_{ij}\right],  t_{ij}= Tr \left [ \rho _{BC} ^{\alpha}(\sigma_{i}\otimes \sigma_{j}) \right ] . 
\end{eqnarray}
These expressions for Eqs. \eqref{Sab}, \eqref{Sac} and \eqref{Sbc} simplify to:

\begin{eqnarray}\label{Sab1}
S_{AB}= 8\rho _{33}^{\alpha}\rho _{55}^{\alpha}+(\rho _{22}^{\alpha}- \rho _{33}^{\alpha} - \rho _{55}^{\alpha})^{2}, \nonumber \\
\end{eqnarray}

\begin{eqnarray}\label{Sac1}
   S_{AC} = 8\rho _{22}^{\alpha}\rho _{55}^{\alpha}+(\rho _{22}^{\alpha}- \rho _{33}^{\alpha} + \rho _{55}^{\alpha})^{2} , \nonumber \\
\end{eqnarray}

\begin{eqnarray}\label{Sbc1}
   S_{BC} = 
  8\rho _{33}^{\alpha}\rho _{22}^{\alpha}+(\rho _{22}^{\alpha} + \rho _{33}^{\alpha} - \rho _{55}^{\alpha})^{2} .\nonumber \\
\end{eqnarray}
As\\
 $\rho _{55}^{\alpha}= P_{\alpha e}(t)= \left |\Bar{U}_{\alpha e }(t) \right |^{2}$,\\
 $\rho _{33}^{\alpha}= P_{\alpha \mu}(t)= \left |\Bar{U}_{\alpha \mu }(t) \right |^{2}$,\\
 $\rho _{22}^{\alpha}= P_{\alpha \tau}(t)= \left |\Bar{U}_{\alpha \tau }(t) \right |^{2}$ are the associated probabilities. Therefore, \eqref{Sab1}, \eqref{Sac1}, \eqref{Sbc1} can be expressed in terms of neutrino oscillation probabilities as
 
 \begin{eqnarray}
S_{AB} = 8 P_{\alpha e}(t) P_{\alpha \mu}(t) + \left\{ P_{\alpha \tau}(t) - P_{\alpha e}(t) - P_{\alpha \mu}(t)\right\}^{2},\nonumber \\
 \end{eqnarray}

\begin{eqnarray}
S_{AC} = 8 P_{\alpha \tau}(t) P_{\alpha e}(t) + \left\{ P_{\alpha \tau}(t) + P_{\alpha e}(t) - P_{\alpha \mu}(t)\right\} ^{2},\nonumber \\
\end{eqnarray}

\begin{eqnarray} 
S_{BC} = 8 P_{\alpha \tau}(t) P_{\alpha \mu}(t) + \left\{ P_{\alpha \tau}(t) - P_{\alpha e}(t) + P_{\alpha \mu}(t)\right\} ^{2}. \nonumber \\
 \end{eqnarray}

The  maximum steering inequality
violation Eq. \eqref{sabc}, $S^{max}(\rho_{ABC})$ can be expressed in terms of neutrino oscillations probabilities as follows:

\begin{align}\label{3f}
S^{max} = \max \left\{ 
8 P_{\alpha e}(t) P_{\alpha \mu}(t) + \left[ P_{\alpha \tau}(t) - P_{\alpha e}(t) - P_{\alpha \mu}(t) \right]^2, \right. & \nonumber \\
8 P_{\alpha \tau}(t) P_{\alpha e}(t) + \left[ P_{\alpha \tau}(t) + P_{\alpha e}(t) - P_{\alpha \mu}(t) \right]^2, & \nonumber \\
\left. 8 P_{\alpha \tau}(t) P_{\alpha \mu}(t) + \left[ P_{\alpha \tau}(t) - P_{\alpha e}(t) + P_{\alpha \mu}(t) \right]^2 \right\} &
\end{align}

\begin{figure*}[ht!]
\centering

\includegraphics[width=75mm]{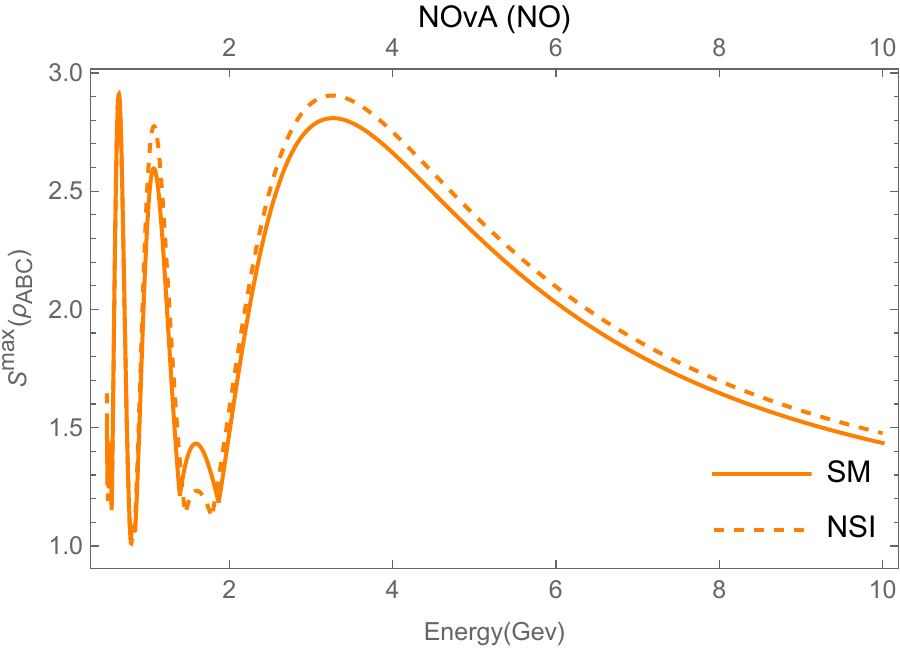}
\includegraphics[width=75mm]{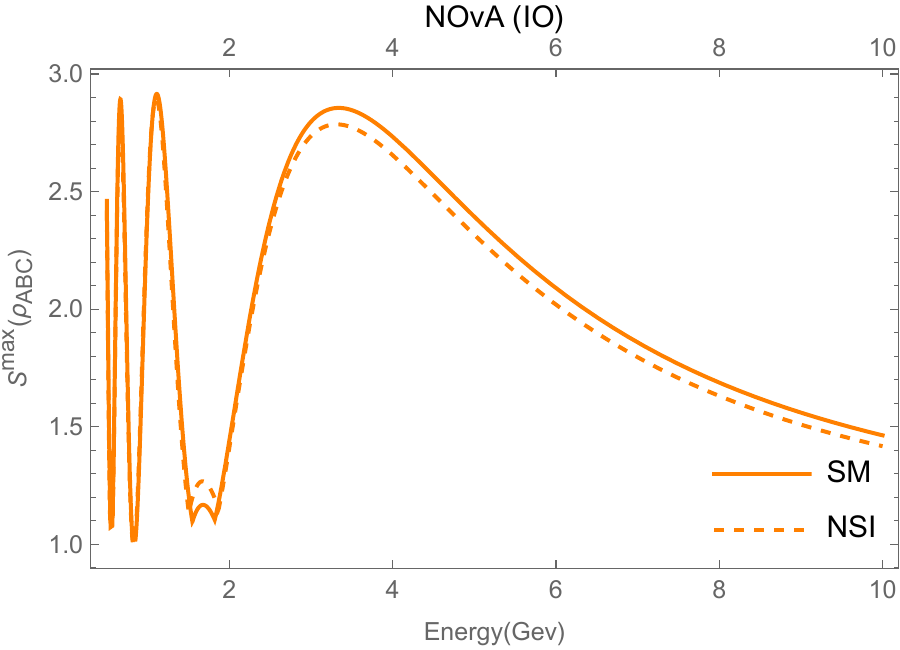}
\\
\includegraphics[width=75mm]{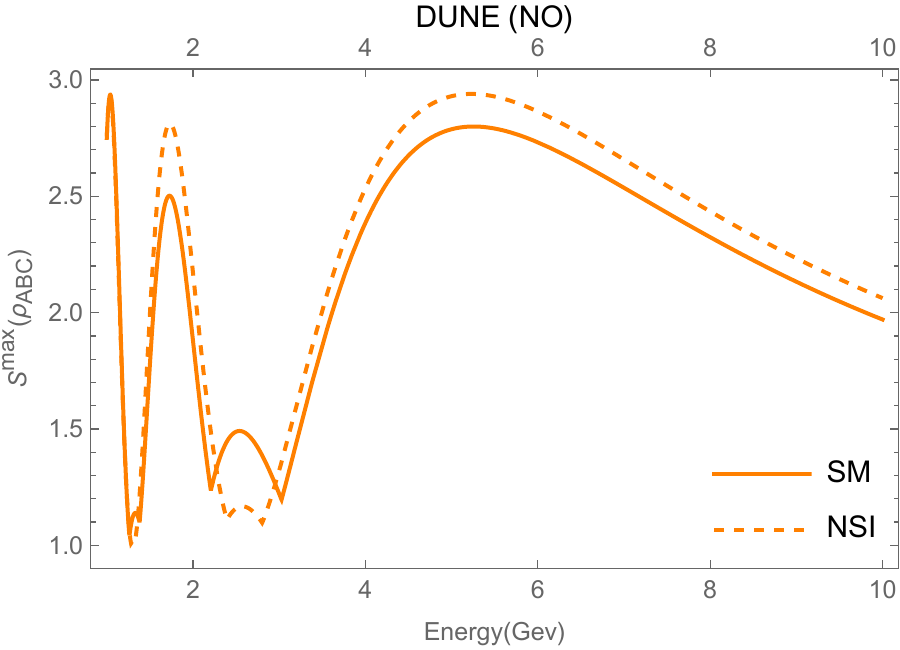}
\includegraphics[width=75mm]{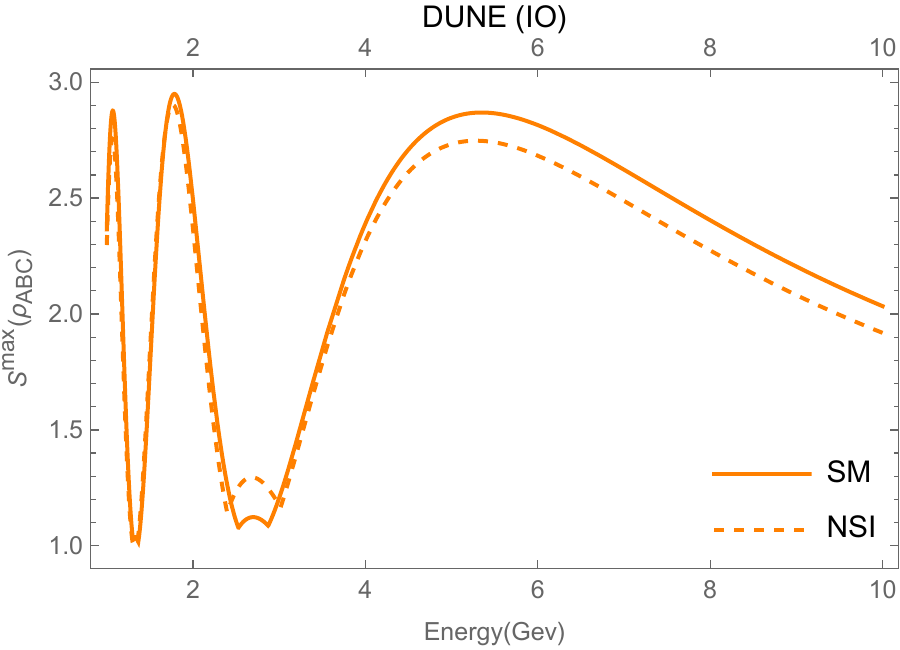}
\caption{The maximum of the steering is shown as a function of energy for the NO$\nu$A (upper panel) and DUNE (lower panel) experiments for the three flavor framework. The solid orange curves represent the SM, while the dashed orange curves indicate NSI for normal ordering (NO) on the left and inverted ordering (IO) on the right.}
\label{fig1}
\end{figure*}
This relation is established for vacuum neutrino oscillations. Moreover, it is also applicable when analyzing probabilities associated with NSI.

If we consider $P_{\alpha \tau}(t)$ is zero i.e., reducing three flavors into two flavor neutrino oscillations, the Eq. \eqref{3f} reduces into Eq. \ref{2f} \cite{Ming:2020}, i.e.,
\begin{eqnarray}\label{pf}
S_{AB}= \left\{P_{\alpha e}(t)+P_{\alpha \mu}(t)\right\}^{2} +8 P_{\alpha e}(t) P_{\alpha \mu}(t).
\end{eqnarray}

\subsection{The relation between $S_{AB}, (S_{BC}, S_{AC})$ and concurrence
$C_{AB} (C_{BC}, C_{AC})$}
For pure bipartite states, the relation is \eqref{2f} 
\begin{eqnarray}
  S_{AB} = 1 + 2 C^{2}_{AB}.
\end{eqnarray}
Where $C_{AB}=2 \sqrt{P_{\alpha e}P_{\alpha \mu}}$ is the concurrence of the state $\rho_{AB}$, indicates that in pure states, greater entanglement corresponds to a more significant violation of the steering inequality.

For tripartite states, the relations between $S_{AB}$, $S_{BC}$, and $ S_{AC}$ with the reduced concurrences are as follows
\begin{eqnarray}
  S_{AB} = 1 + 2 C^{2}_{AB} -C^{2}_{AC}-C^{2}_{BC} ,
\end{eqnarray}
\begin{eqnarray}
  S_{AC} = 1 + 2 C^{2}_{AC} -C^{2}_{AB}-C^{2}_{BC},
\end{eqnarray}
\begin{eqnarray}
  S_{BC} = 1 + 2 C^{2}_{BC}-C^{2}_{AB}-C^{2}_{AC}.
\end{eqnarray}
Where $C_{AB}=2\sqrt{P_{\alpha e}P_{\alpha \mu}}, C_{AC}=2 \sqrt{P_{\alpha e}P_{\alpha \tau}}$, and $C_{BC}=2 \sqrt{P_{\alpha \mu}P_{\alpha \tau}}$ are the reduced concurrence of the state $\rho_{ABC}$.

\begin{table}[htb!] 
\centering
    \caption{Standard neutrino oscillation parameters \cite{Esteban:2024}}
 \label{Tab1}
    \begin{tabular}{|c | c| c|}
    \hline
    \hline
     Input Parameters & Ordering & Best fit$\pm 1\sigma$\\
    \hline
    $\Delta m_{21}^2 / 10^{-5} \, \text{eV}^2$ & NO & $7.49_{-0.19}^{+0.19}$ \\
    & IO & $7.49_{-0.19}^{+0.19}$ \\
    \hline
    $\sin^2{\theta_{12}}$ & NO & $0.307_{-0.011}^{+0.012}$ \\
    & IO & $0.308_{-0.011}^{+0.012}$ \\
    \hline
    $\Delta m_{31}^2 / 10^{-3} \, \text{eV}^2$ & NO & $2.534_{-0.023}^{+0.025}$ \\
    & IO & $-2.510_{-0.023}^{+0.025}$ \\
    \hline
    $\sin^2{\theta_{23}}$ & NO & $0.561_{-0.015}^{+0.012}$ \\
    & IO & $0.562_{-0.015}^{+0.012}$ \\
    \hline
    $\sin^2{\theta_{13}}$ & NO & $0.02195_{-0.00058}^{+0.00054}$ \\
    & IO & $0.02224_{-0.00057}^{+0.00056}$ \\
    \hline
    $\delta/^{\circ}$ & NO & $177_{-20}^{+19}$ \\
    & IO & $285_{-28}^{+25}$ \\
    \hline
    \end{tabular}
\label{Inputs}
\end{table}

\begin{table}[t]
	\centering
	\caption{The NSI parameters as presented in \cite{Coloma2023} are considered to be real-valued, and their 2$\sigma$ upper limits have been used to produce the NSI plots.}
	\label{Tab2}
	\begin{tabular}{|c|c|c|}
		\hline
        NSI Parameters  & Range ($1 \sigma$)& Range ($2 \sigma$)\\      
		\hline\hline
				$\epsilon_{ee}^{\oplus}$ & [-0.30, 0.20] $\oplus$ [0.95, 1.3] & [-1.00, 1.4]\\
		\hline
		$\epsilon_{e \mu}^{\oplus}$ &[-0.12, 0.011]& [-0.20, 0.09] \\
		\hline
		$\epsilon_{e \tau}^{\oplus}$ & [-0.16, 0.083]& [-0.24, 0.30]\\
		\hline
		$\epsilon_{\mu \mu}^{\oplus}$ & [-0.43, 0.14] $\oplus$ [0.91, 1.3]& [-0.80, 1.4] \\
		\hline
		$\epsilon_{\mu \tau}^{\oplus}$ & [-0.047, 0.012]& [-0.021,0.021] \\
		\hline
		$\epsilon_{\tau \tau}^{\oplus}$ &[-0.43, 0.21] $\oplus$ [0.83, 1.3]& [-0.85, 1.4]\\
		\hline
	\end{tabular}
\end{table}

\section{Results and Discussions}
\label{sec5}

In this section, we study the effect of NSI on the steering measure in the three-flavor neutrino oscillations framework, both in the SM and the presence of NSI. The analysis includes two experimental setups: the ongoing NO$\nu$A experiment and the upcoming DUNE experiment. We investigate how NSI modify the steering measure in neutrino oscillations in the context of both experimental set-ups and compare the results with SM predictions to highlight the impact of NSI on steering.

\textbf{NO$\nu$A} (NuMI Off-Axis $\nu_{e}$ Appearance) \cite{Adamson:2016} is a prominent long-baseline accelerator neutrino oscillations experiment has a baseline of L = 810 km.  Based at Fermilab, NO$\nu$A uses a beam of muon neutrinos produced at the NuMI (Neutrinos at the Main Injector) facility. Key objectives of NO$\nu$A include probing the neutrino mass hierarchy, investigating CP violation in the lepton sector, and refining measurements of oscillation parameters such as $\theta_{23}$ and $\delta_{CP}$. 

\textbf{DUNE} (Deep Underground Neutrino Experiment) \cite{DUNE:2015lol, Cherchiglia:2024, Abbaslu:2024} is a next-generation, long-baseline neutrino experiment of 1300 km baseline under construction, designed to study neutrino oscillations with a focus on resolving key unknowns in neutrino physics, including $\delta_{CP}$, the neutrino mass ordering, and the octant of $\theta_{23}$. DUNE is expected to have unprecedented sensitivity to oscillation parameters, making it ideal for investigating NSI effects alongside SM interactions.

Table \ref{Tab1} presents the standard neutrino oscillation parameters, and Table \ref{Tab2} lists the NSI parameters within the context of three-flavor neutrino oscillations.

In Figure \ref{fig1}, the steering inequality is studied in both the SM and NSI scenarios, exhibiting different behaviour across various energy ranges. For the \textbf{NO$\nu$A} experiment within the SM, the maximum steering value reaches 2.9 for both NO and IO but at different energy values: 0.8 GeV for NO and 1.26 GeV for IO. At E approx 1.26 GeV and in E $\geq$ 2 GeV, NO shows enhancement of the NSI compared to SM, with enhancement of approx 6$\%$ at E $\approx$ 1.26 GeV and approx 3.5$\%$ at E $\approx $ 3.5GeV, while in IO, the NSI exhibits suppression over SM in the higher energy region. In the energy range of [1.54 - 2] GeV, SM exceeds the NSI value, which is inconsistent with the behaviour observed in other energy regions for NO, where there is an enhancement of approx 11.4$\%$.

In the \textbf{DUNE} experiment under the SM framework, the maximum steering value attains 2.9 at E approx 1.22 GeV for NO, whereas for IO, it reaches a higher peak of 2.95 at E $  \approx $ 2 GeV.

\begin{figure}[ht!]
\centering

\includegraphics[width=75mm]{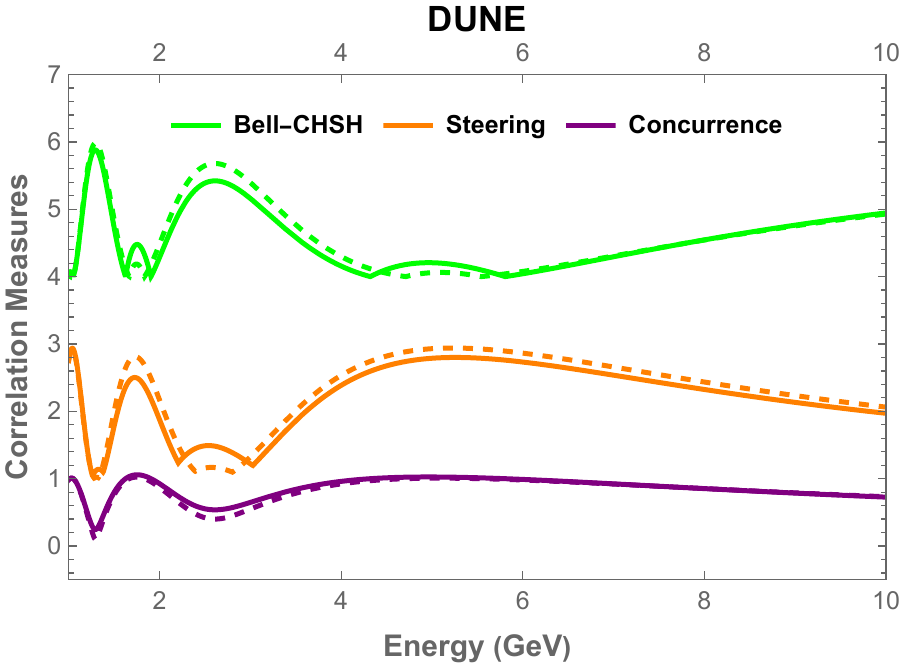}

\caption{The plot illustrates the behavior of nonlocality, steering, and entanglement in the DUNE experimental setup for NO, considering both the SM and NSI. Nonlocality is quantified using the Bell-CHSH inequality, quantum steering is measured through steering inequality, and entanglement is characterized by concurrence. The solid lines correspond to the SM predictions, while the dashed lines reflect the predictions that include NSI. The color scheme is as follows: green represents Bell-CHSH inequality, orange denotes steering inequality, and purple indicates concurrence.}
\label{fig2}
\end{figure}
In the low-energy range of approx [1.5 - 2] GeV for NO, NSI effects dominate, with the steering value for NSI being $\approx$ 12$\%$ enhancement that of the SM. This indicates that NSI contributions are more pronounced and could potentially influence the oscillation pattern more strongly at these lower energies. In the intermediate energy range of $\approx$ [2.3 - 3.1] GeV, the trend reverses, with the SM steering value surpassing that of the NSI by about 21$\%$ for NO. For the IO, the NSI steering value exceeds that of the SM by approx 15$\%$ at the energy $\approx$ [2.5 - 3.1] GeV. This occurs around the energy where the DUNE experiment observes maximum flux for both NO and IO. In the higher energy range of $\geq$ 3 GeV in NO, the influence of NSI once again surpasses the SM value by approximately 5\%, indicating a resurgence of NSI effects at higher energies. However, in the case of inverted ordering, this trend is reversed.

Figure \ref{fig1} presents the steering measure $S_{max}(\rho_{ABC} )$ as a function of neutrino energy for both the NO$\nu$A and DUNE experiments under the SM and NSI frameworks. The results indicate that quantum steering is consistently observed across the analyzed energy ranges, with $S_{max}$ exceeding 1 in all cases. This implies that the neutrino states involved in the NO$\nu$A and DUNE experiments are indeed steerable, suggesting that neutrino oscillation measurements cannot be explained by local hidden variable theory, thereby reinforcing the quantum nature of neutrinos.

Steering offers a complementary perspective on the role of quantum correlations in neutrino oscillations, as the steering measure is directly expressed in terms of neutrino oscillation probabilities. Since our study focuses on the effects of NSI on steering, any modifications that NSI induces in oscillation probabilities will also be reflected in the steering measure. Thus, by detecting NSI effects through probability measurements, we can reinforce their presence through steering observations. This dual approach enhances our understanding of NSI phenomena and strengthens our ability to quantify their impact on neutrino dynamics.

Figure \ref{fig2} illustrates a comparative analysis of three quantum correlation measures: nonlocality, steering, and entanglement within the context of the DUNE experiment. Here, nonlocality is quantified using the Bell-CHSH inequality \cite{Qin:2015,Li:2022}, steering is assessed through steering inequality, and entanglement is evaluated via concurrence \cite{Wootters1998, Guo2019, Ming2021}. The plot investigates the influence of these quantum correlations on the three-flavor neutrino oscillation framework, providing a detailed comparison of results under both the SM and NSI scenarios. Quantum steering represents a quantum correlation that sits between Bell nonlocality and entanglement \cite{Quintino:2015}. In this analysis, we seek to explore this property within the context of the neutrino system. A distinct hierarchy emerges among these quantum correlations: states that demonstrate Bell nonlocality are inherently steerable and entangled.  However, not every entangled state is steerable, nor does it necessarily violate the Bell inequality. 

\section{Conclusion}
\label{sec6}
In this work, we examine the impact of NSI on steering for the three-flavor neutrino oscillations, considering NO$\nu$A and DUNE experimental setups. DUNE, having the longer baseline, demonstrates a more significant deviation for steering toward NSI than NO$\nu$A. For DUNE, in the energy range of [2.3 - 3.1] GeV, the steering value shows a deviation of about 21\% from the SM to NSI for NO, while in the case of IO, the steering value increases by approximately 15\% compared to the SM. We also compared nonlocality, steering, and entanglement, where the Bell-CHSH captures nonlocality, steering measures how steerable the neutrino state is, and concurrence captures entanglement. We have also expressed steering in terms of three-flavor neutrino oscillation probabilities for vacuum and NSI. Furthermore, we explore the relation between steering inequality and concurrence in neutrinos. 

\section{Acknowledgement}
The authors express sincere gratitude to Prof. Rukmani Mohanta for invaluable support and comments.

\end{document}